\documentclass[
 reprint,
 amsmath,amssymb,superscriptaddress,
 aps,
 prl,
groupedaddress
]{revtex4-1}

\usepackage{graphicx}
\usepackage{dcolumn}
\usepackage{bm}
\usepackage{color}
\usepackage{ulem}
\usepackage{float}
\usepackage{textcomp}
\restylefloat{table}
\usepackage{verbatim}
\usepackage{epstopdf}
\usepackage[mediumspace,mediumqspace,Grey,squaren]{SIunits}

\begin{document}

\title{Reorientation of the stripe Phase of 2D Electrons by a Minute Density Modulation}
\date{\today}

\author{M. A.\ Mueed}
\author{Md. Shafayat\ Hossain}
\author{L. N.\ Pfeiffer}
\author{K. W.\ West}
\author{K. W.\ Baldwin}
\author{M.\ Shayegan}
\affiliation{Department of Electrical Engineering, Princeton University, Princeton, New Jersey 08544, USA}

\begin{abstract}

Interacting two-dimensional electrons confined in a GaAs quantum well exhibit isotropic transport when the Fermi level resides in the first excited ($N=1$) Landau level. Adding an in-plane magnetic field ($B_{||}$) typically leads to an anisotropic, stripe-like (nematic) phase of electrons with the stripes oriented perpendicular to the $B_{||}$ direction. Our experimental data reveal how a periodic density modulation, induced by a surface strain grating from strips of negative electron-beam resist, competes against the $B_{||}$-induced orientational order of the stripe phase. Even a minute ($<0.25\%$) density modulation is sufficient to reorient the stripes along the direction of the surface grating.

\end{abstract} 

\maketitle

Strongly correlated electronic systems host a spectacular array of quantum phases of matter. Amongst these, the stripe phase (SP), associated with a spontaneous symmetry breaking, has drawn considerable attention over the years \cite{Fradkin.ARCMP.2010}. In this many-body ground state, electrons arrange themselves in periodic clusters to form stripes that resemble the classical nematic liquid crystals of elongated molecules \cite{deGennes.1993,Chaikin.1995}. A readily observable signature of a SP is the dramatic anisotropy in the resistivity along and perpendicular to the stripes' direction. Transport measurements have confirmed such anisotropy in a variety of systems including the very high-quality two-dimensional electron systems (2DESs) in modulation doped semiconductors under perpendicular magnetic fields ($B_{\perp}$) \cite{Cooper.SSC.2001,Willett1.PRL.2001,Endo.PRB.2002,Fradkin.PRB.1999,Willett.PRL.2001,Koulakov.PRL.1996,Fogler.PRB.1996,Moessner.PRB.1996,Pan.PRL.1999,Lilly1.PRL.1999,Lilly.PRL.1999,Cooper.PRB.1999,Du.SSC.1999,Shayegan.physica.2000}, strontium ruthenate materials \cite{Borzi.Science.2007} and high-$T_c$ superconductors \cite{Ando.PRL.2002,Hinkov.Science.2008}. Although widely studied and the subject of intense renewed interest \cite{Cooper.SSC.2001,Willett1.PRL.2001,Endo.PRB.2002,Willett.PRL.2001,Friess.PRL.2014,Fradkin.PRB.1999,Pan.PRL.1999,Lilly1.PRL.1999,Lilly.PRL.1999,Cooper.PRB.1999,Du.SSC.1999,Shayegan.physica.2000,Koulakov.PRL.1996,Fogler.PRB.1996,Moessner.PRB.1996,Fradkin.PRB.1999,Sambandamurthy.PRL.2008,Zhu.PRL.2009,Zhang.PRL.2010,Shi2.PRB.2015,Shi1.PRB.2016,Xia.PRL.2010,Liu1.PRB.2013,Koduvayur.PRL.2013,Liu.PRB.2013,pollanen.PRB.2015,Samkharadze.NatPhy.2016,Shi.PRB.2016}, microscopic details of the 2DES SP still remain elusive. A main reason is that techniques such as scanning tunneling microscopy, which are commonly used to probe the electrons' spatial configuration, are not suitable in this case because the 2DES is typically buried deep under the surface, thus limiting the achievable resolution. In this manuscript, we report a new method for studying the structure and energetics of the SP by perturbing the 2DES with an external, gentle periodic density modulation.


In its higher ($N\geq2$) Landau levels (LLs), a 2DES prefers a SP as its ground state at the half-fillings \cite{Koulakov.PRL.1996,Fogler.PRB.1996,Moessner.PRB.1996,Fradkin.PRB.1999,Jungwirth.PRB.1999,Rezayi.PRL.2000,Fradkin.ARCMP.2010,Lilly1.PRL.1999,Du.SSC.1999}. Hartree-Fock calculations predict that in such a phase, the 2DES becomes density modulated in parallel stripes of the neighboring integer quantum Hall states \cite{Koulakov.PRL.1996,Fogler.PRB.1996,Moessner.PRB.1996,Fradkin.PRB.1999}. For example, the spatial topography of an ideal SP at $\nu=9/2$ would consist of alternating stripes of $\nu=4$ and 5. Because of disorder, and at finite temperature, the SP is more likely a nematic phase in real samples; nevertheless for simplicity, we refer to this phase as a SP. In such a SP, the resistivity becomes anisotropic: it is large when current passes perpendicular to the quantum Hall stripes (hard-axis) and small when it passes parallel to them (easy-axis). In contrast to the $N\geq2$ LLs, in the first excited ($N=1$) LL, i.e. at filling factors $\nu=7/2$ and 5/2, there is a close competition between the anisotropic SP and the isotropic fractional quantum Hall state \cite{Jungwirth.PRB.1999,Rezayi.PRL.2000}. Generally, the latter is preferred in the absence of any in-plane field ($B_{||}$). However, if $B_{||}$ is applied, the system typically favors a SP with its hard-axis oriented $parallel$ to $B_{||}$ \cite{Cooper.PRB.1999,Lilly.PRL.1999,Pan.PRL.1999,Jungwirth.PRB.1999,Rezayi.PRL.2000}. 
The $B_{||}$-induced SPs are more robust than those without $B_{||}$ which usually do not survive at $T\gtrsim100$ mK \cite{Fradkin.ARCMP.2010}.

\begin{figure}
\includegraphics[width=.45\textwidth]{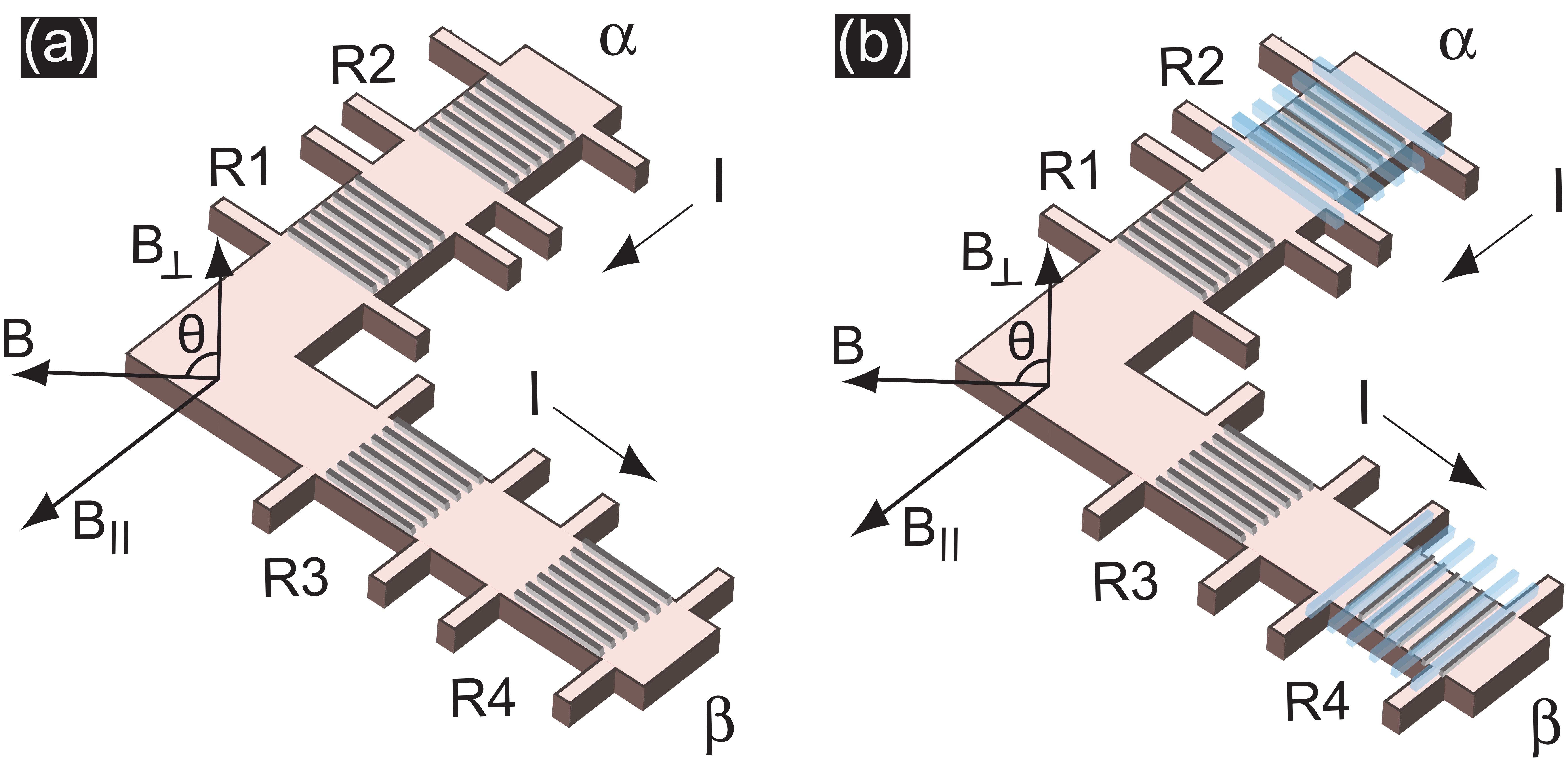}
\caption{\label{fig:Fig1} (color online) L-shaped Hall bar with arms $\alpha$ and $\beta$ is designed to probe the resistivity along the perpendicular and parallel directions to the SP. The sample is rotated around the axis of arm $\beta$ with $\theta$ denoting the tilt angle. $B_{||}$, directed along arm $\alpha$, induces stripes (shown as dark gray) perpendicular to its direction. In (a) we show the sample without any imposed density modulation. Although the SP is represented as gray lines only in the different regions of the Hall bar (R1, R2, R3, and R4), it exists throughout the whole Hall bar. As illustrated in (b) a surface superlattice of negative electron beam resist (blue lines) is fabricated perpendicular to the current direction in R2 and R4 to induce a small density modulation. We observe experimentally a reorientation of the stripes in R4, as shown by the 90$^\circ$ rotated gray lines.}
\end{figure}


\begin{figure*}
\includegraphics[width=.98\textwidth]{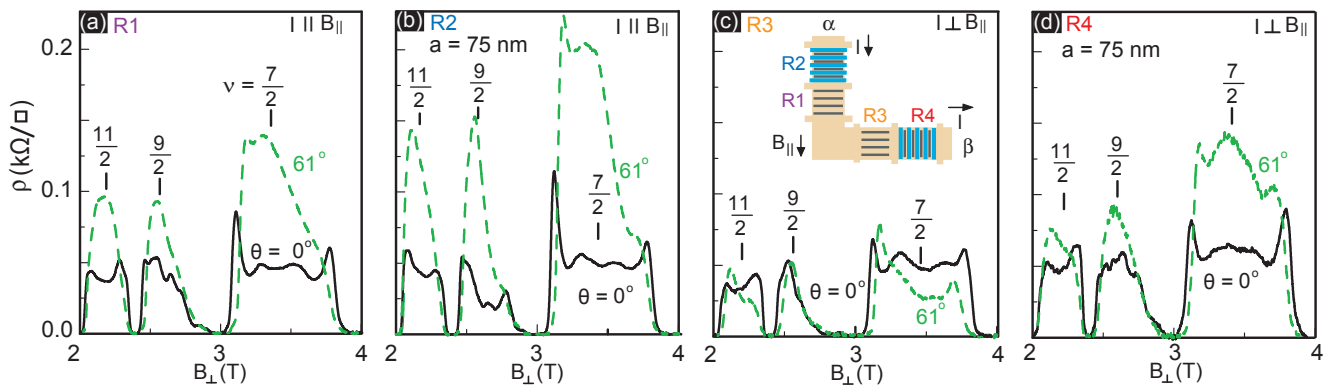}
\caption{\label{fig:Fig1} (color online) Magnetoresistivity traces at $T\simeq0.1$ K for all four regions of a sample with $a=75$ nm. In each panel, we mark the positions of $\nu=11/2$, 9/2 and 7/2. The solid black traces are for $\theta=0^\circ$ while the dashed green traces for $\theta=61^\circ$. In Fig. 1(c), we also show an L-shape Hall bar with the regions labeled in different colors. Note that the same color code is used for the panel titles.}
\end{figure*}

Here, we report that an imposed periodic density modulation with a $minute$ amplitude can reorient the $B_{||}$-induced SP. This is remarkable since the SP is expected to have a much larger modulation. Our data provide quantitative evidence that the $parallel$ and $perpendicular$ orientations of the SP with respect to the $B_{||}$-direction must be energetically very close, corroborating theoretical predictions which have not been confirmed thus far \cite{Jungwirth.PRB.1999}. Figure 1 captures our approach. We use an L-shaped Hall bar with four regions marked as R1, R2, R3, and R4. When a sufficiently large $B_{||}$ is introduced by tilting the sample in magnetic field, a SP becomes the ground state at $\nu=7/2$. Since the stripes are oriented perpendicular to $B_{||}$, current will flow perpendicular to the stripes in arm $\alpha$,  making R1 and R2 the hard-axis. In contrast, R3 and R4 are both the easy-axis as current passes parallel to the stripes in arm $\beta$. In Fig. 1(b), we illustrate the concept of introducing an additional periodic density modulation in R2 and R4. The blue lines which represent the external modulation are oriented perpendicular to the current. This orientation is $parallel$ to the $B_{||}$-induced stripes in R2 and $perpendicular$ in R4. 
Our data, taken in the configuration of Fig. 1(b), show that R1 and R3 are hard- and easy-axis, as expected. R2 remains a hard-axis whereas R4 surprisingly transforms into a hard-axis. In Fig. 1(b), we show R4's change from easy- to hard-axis, i.e. a \textit{reorientation} of the $B_{||}$-induced SP, by rotating the gray lines by $90^\circ$.

Our sample, grown via molecular beam epitaxy, is a 30-nm-wide, GaAs (001) quantum well (QW) located 135 nm under the surface. The QW is flanked on each side by 95-nm-thick Al$_{0.24}$Ga$_{0.76}$As spacer layers and Si $\delta$-doped layers. The 2DES density is $n=2.95\times10^{11}$ cm$^{-2}$, and its low temperature mobility is $\mu\sim 20\times$$10^{6}$ cm$^{2}$/Vs. Each region of the L-shaped Hall bar has a length of 100 $\micro$m and width of 50 $\micro$m. As shown in Fig. 1(b), we pattern the surfaces of R2 and R4 with a strain-inducing superlattice of period $a$. We study several samples with $a$ = 75, 100 and 175 nm. The superlattice, made of negative electron-beam resist, imparts a density modulation of the same period to the 2DES through the piezoelectric effect in GaAs \cite{Endo.PRB.2005,Cusco.Surf.1994,Kamburov.PRB.2012,Endo.PRB.2000,Skuras.APL.1997,Long.PRB.1999,Endo.PRB.2005}. 
The modulation amplitude will be discussed later in the paper. Measurements were mostly carried out in a dilution refrigerator; our best estimate for the lowest electron temperature is about 0.1 K.


Figure 2 presents the data for a sample with $a=75$ nm. The black and green traces are taken at $\theta=0$ and $61^\circ$, respectively. We primarily focus on the longitudinal resistivity ($\rho$) at $\nu=$ 7/2. Note that the current direction is perpendicular to the expected $B_{||}$-induced stripes in R1 and R2 but parallel in R3 and R4. In the presence of $B_{||}$, the resistivity at $\nu=7/2$ increases for R1 and R2 (Figs. 2(a) and (b)) confirming that both regions of arm $\alpha$ are indeed hard-axis for current. At $\theta=61^\circ$, R2 shows a larger $\rho$ compared to R1, implying that the external modulation causes an enhancement of the transport anisotropy. In contrast, the resistivity near $\nu=7/2$ decreases for R3 (see Fig. 2(c)) revealing that, as expected, this is an easy-axis. However, instead of becoming an easy-axis, R4 transforms into a hard-axis as evidenced by the increase of the $\nu=7/2$ resistivity in the green trace of Fig. 2(d). This transport behavior points to a reorientation of the $B_{||}$-induced stripes in R4.

\begin{figure}
\includegraphics[width=.48\textwidth]{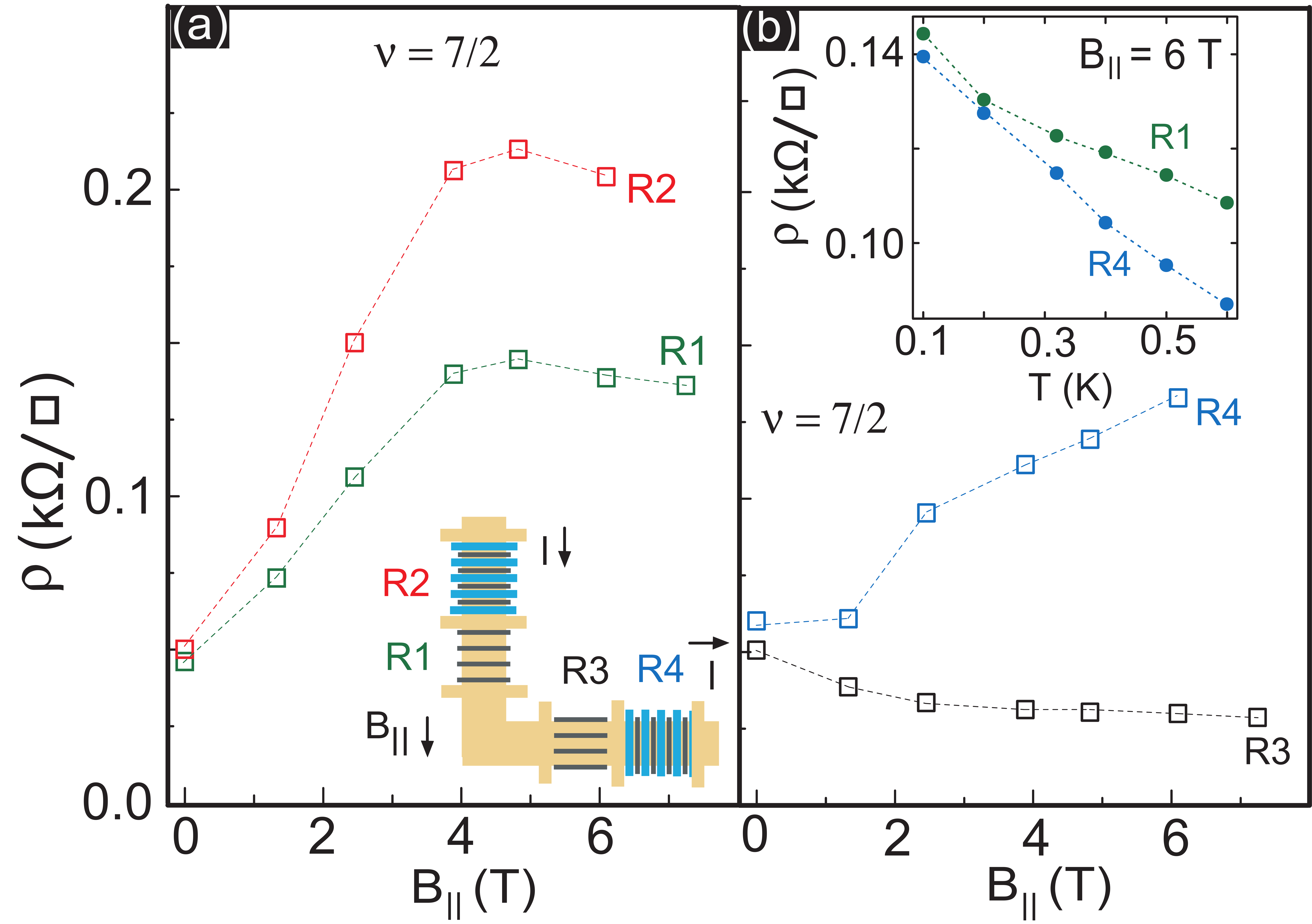}
\caption{\label{fig:Fig1} (color online) The $\nu=7/2$ resistivity, at $T\simeq0.1$ K, is plotted as a function of $B_{||}$ for different regions: (a) R1 and R2; (b) R3 and R4. Inset: Temperature dependence of the $\nu=7/2$ resistivity for R1 and R4 at $B_{||}=6$ T.}
\end{figure}

\begin{figure}
\includegraphics[width=.3\textwidth]{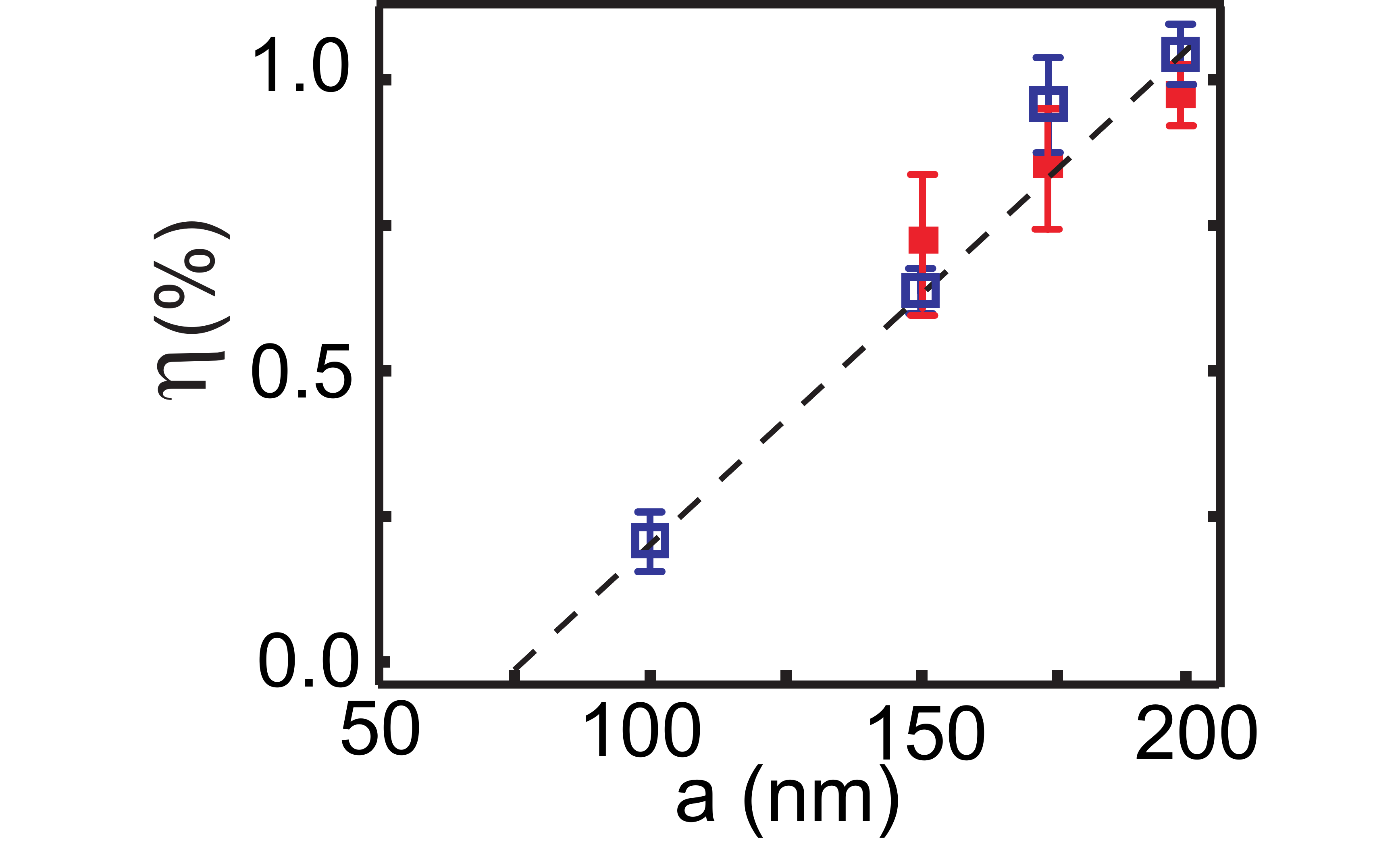}
\caption{\label{fig:Fig1} (color online) Amplitude of the external modulation $\eta$ versus the modulation period $a$. The red and blue squares represent values of $\eta$ estimated from the fit of COs and positions of extinction field, respectively (see Supplemental Material for details). Note that the dashed line extrapolates to $\eta\simeq$ 0 for $a=75$ nm.}
\end{figure}

To better evaluate the role of the external density modulation, in Fig. 3 we show plots of $\rho$ at $\nu=7/2$ for all regions as a function of $B_{||}$. Figure 3(a) illustrates that for both R1 and R2, $\rho$ increases as a function of $B_{||}$. However, the relative increase is clearly larger for R2 implying that the external modulation effectively enhances the strength of the SP in R2. Focusing on R3 and R4, as shown in Fig. 3(b), the $\nu=7/2$ resistivity decreases for R3 as a function of $B_{||}$. The resistivity ratio between the unpatterned regions R1 and R3, i.e. hard- and easy-axis, reaches a maximum of $\sim5.5$ at $B_{||}=4.8$ T. Note that van der Pauw (vdP) samples, more commonly used to study the SPs, show significantly larger anisotropy than Hall bar samples \cite{Lilly1.PRL.1999}. The exponentially decaying current density between the current and voltage contact pairs in a vdP geometry causes the easy-axis resistance to be very small and thus exaggerates the anisotropy \cite{Simon.PRL.1999}. We emphasize that an anisotropy of $\sim5.5$ in a Hall bar sample, if translated into vdP geometry, converts to $\sim40$ which is comparable to the typical large values observed in vdP samples \cite{Simon.PRL.1999}. Figure 3(b) also contains the data for R4. Unlike in R3, the resistivity $increases$ as a function of $B_{||}$. This increase, resembling that of R1 and R2, provides compelling evidence that the external modulation causes the $B_{||}$-induced stripes to rotate by $90^\circ$ in R4. 

Before a detailed discussion, we point to two features of Figs. 2 and 3 data. First, the $B_{||}=0$ resistivities of all the patterned and unpatterned regions are similar, suggesting that the external modulation is merely perturbative at $B_{||}=0$ and does not directly influence the transport. This is consistent with the very small amplitude of the density modulation (see next paragraph). Second, to compare the robustness of the $intrinsic$ and $reoriented$ hard-axis, in Fig. 3(b) inset we show the temperature dependence of the resistivities for R1 and R4 at $B_{||}=6$ T. The values of $\rho$ are close at low-$T$ but decrease more rapidly for R4 than R1 as $T$ is increased. Moreover, when normalized to the $B_{||}=0$ value, $\rho$ approaches unity for R4 at $T\gtrsim0.5$ K, whereas it still stays $\simeq1.35$ for R1. This indicates that the hard-axis in R4 is weaker suggesting that the reorientation phenomenon induced by the external modulation is fragile.

In Fig. 4, we plot the amplitude of the external modulation $\eta$ stemming from the surface superlattice as a function of the modulation period $a$; $\eta$ is measured as a ratio to the Fermi energy ($E_F$). When subjected to a periodic external modulation, the magnetoresistivity of a 2DES shows commensurability oscillations (COs) at small $B_{\perp}$ \cite{Mirlin.PRB.1998,Peeters.PRB.1992,Skuras.APL.1997,Long.PRB.1999,Endo.PRB.2005,Cusco.Surf.1994,Endo.PRB.2000,Kamburov.PRB.2012,Gerhardts.PRL.1989,Winkler.PRL.1989,Weiss.Europhys.1989,Beenakkaer.PRL.1989,Beton.PRB.1990}.
Moreover, before the onset of COs, it also shows a positive magnetoresistivity peak known as the extinction field feature \cite{Beton.PRB.1990,Endo.PRB.2000,Endo.PRB.2005,Kamburov.PRB.2012}.
As described in details in the Supplemental Material \cite{footnote3}, we can estimate $\eta$ from COs' amplitude or the extinction field. Figure 4 shows that $\eta$ decreases with decreasing $a$. For shorter periods, as their ratio to the 2DES depth from the surface becomes smaller, the higher harmonics of the potential modulation are attenuated, resulting in a weaker amplitude \cite{Davies.PRB.1994,Long.PRB.1999,Kamburov.PRB.2012,Lu.APL.1994}. For $a=75$ nm, we cannot estimate $\eta$ using the above methods since no COs or extinction field feature are observed (see Supplemental Material). However, from its dependence on $a$, we conclude that $\eta<0.25\%$. Such a weak modulation is consistent with the fact that $a=$ 75 nm is significantly smaller than our sample's 2DES depth (135 nm).

Data presented in Figs. 2-4 provide clear evidence that an extremely small ($<0.25\%$) external modulation is sufficient to reorient the $B_{||}$-induced stripes. Is this reasonable? In an ideal $\nu=7/2$ SP, composed of $\nu=3$ and 4 quantum Hall stripes, the density modulation should follow the filling factor modulation, i.e., $\sim30\%$. However, at $\nu=7/2$, the electron's cyclotron orbit size is about $2l_B$, while the period ($\lambda$) for the $\nu=7/2$ is expected to be $\simeq4.7l_B$ \cite{Fogler.PRB.1996} ($l_B=\hbar/eB_{\perp}$ is the magnetic length). This implies that the electron's orbit barely fits within the width of the stripes ($\lambda/2$). As a result, the density modulation is theoretically expected to be much smaller, only about $20\%$ of the ideal filling factor modulation \cite {Fogler.PRB.1996}. One would therefore expect a density modulation of only $\sim6\%$ for the $\nu=7/2$ SP (and $\sim13\%$ for the SP at $\nu=5/2$). Recent nuclear magnetic resonance measurements have in fact revealed a density modulation of $\sim20\%$ for the $\nu=5/2$ SP \cite{Friess.PRL.2014} which is much smaller than the ideal ($40\%$) value, and closer to the predicted value ($\sim 13\%$). Although no experimental quantification exists for the $B_{||}$-induced $\nu=7/2$ SP, the above analysis clearly indicates that the expected $\sim6\%$ density modulation is substantially larger than the external density modulation we are imposing in our samples ({$<0.25\%$). Yet, surprisingly, we observe a reorientation. This is, however, plausible: our data imply that the SP's $parallel$ and $perpendicular$ orientations must be energetically very close. Indeed, calculations show that, despite possessing a relatively large intrinsic modulation, the energy \textit{difference} between these two orientations is as small as $\sim0.2$ K per electron for the $N=1$ LL at a density similar to our 2DES density \cite{Jungwirth.PRB.1999}. In units of $E_F$, this difference converts to $\eta\sim0.2\%$. Since our estimate for the $a=75$ nm case is $\eta<0.25\%$, it is reasonable that even such a weak external modulation can compensate for the energy difference and cause a reorientation.

\begin{figure}
\includegraphics[width=.484\textwidth]{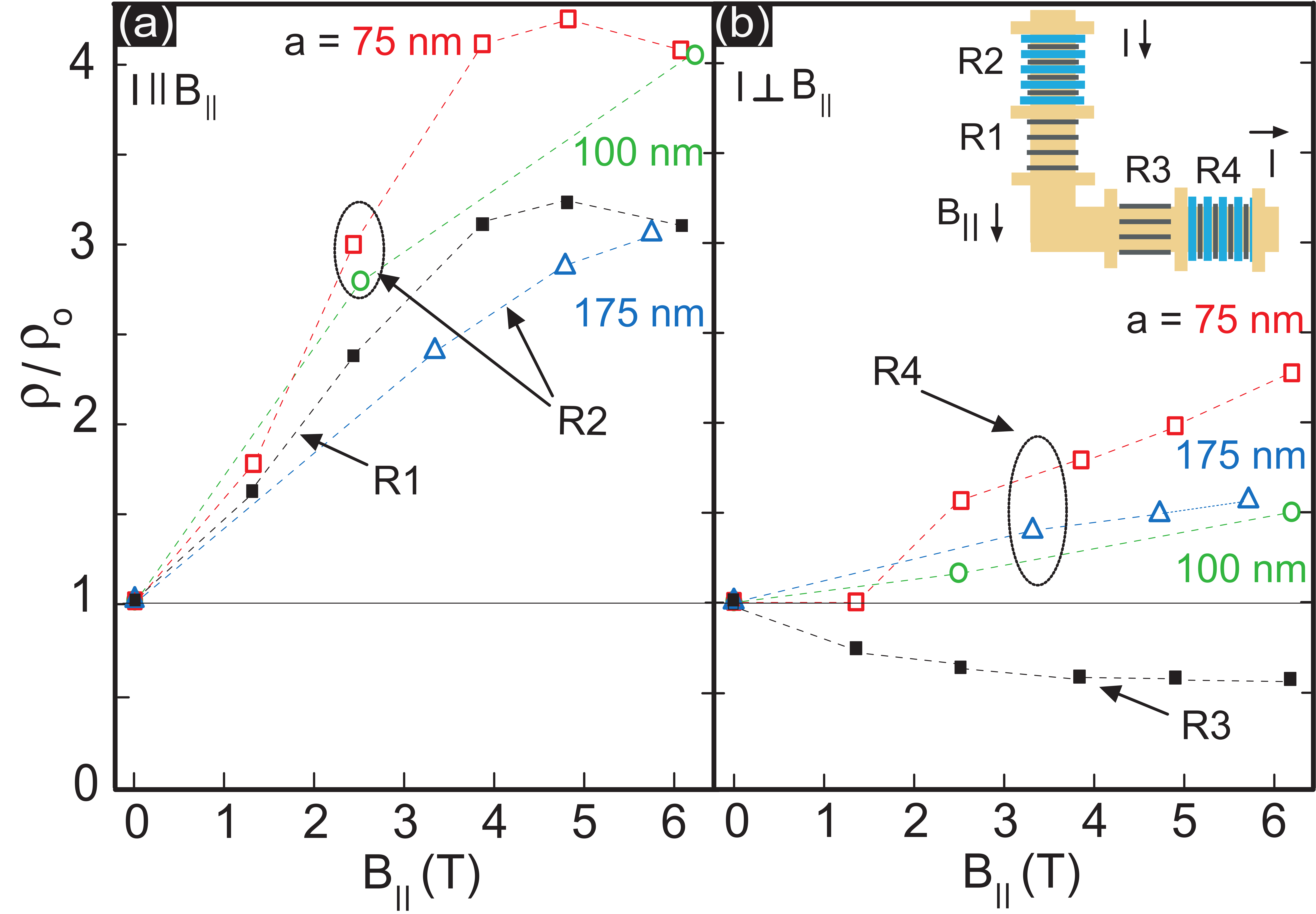}
\caption{\label{fig:Fig1} (color online) The $\nu=7/2$ resistivity, normalized to its $B_{||}=0$ value ($\rho_o$), as a function of $B_{||}$ for the patterned regions: (a) R2, and (b) R4. We show data for the $a$ = 75, 100 and 175 nm cases. For comparison, we also include data (solid squares) for the unpatterned R1 and R3.}
\end{figure}

Finally, we discuss the correlation between the periods of the $B_{||}$-induced stripes and the external modulation. Note that $\lambda\simeq65$ nm for the $\nu=7/2$ SP in our sample ($l_B\simeq13.5$ nm). We studied three samples with $a=$ 75, 100 and 175 nm. 
Figures 5(a) and (b) show data for R2 and R4, respectively; we also include data for the relevant unpatterned regions, R1 and R3. In these figures, we plot $\rho$ at $\nu=7/2$, normalized to the $B_{||}=0$ value, as a function of $B_{||}$. Relative to the unpatterned R1, $\rho$ for the patterned region R2 increases most for $a$ = 75 nm as seen in Fig. 5(a). We expect that in R2, the external modulation would facilitate the $B_{||}$-induced stripes since they are parallel to each other and thus enhance the resistivity \cite{footnote1}. Although, as illustrated in Fig. 4, the modulation for the $a=$ 75 nm superlattice is the weakest, it shows the largest enhancement in Fig. 5(a). This is plausible since $a=75$ nm is comparable to $\lambda$. We would therefore expect that $a=75$ nm should also be the most effective in causing the reorientation. Indeed, Fig. 5(b) data corroborate our reasoning. Although $\rho$ increases for all three periods, signaling SP's reorientation in R4 , the $a=75$ nm data exhibit the highest relative increase despite the weakest modulation. Clearly, as $a$ becomes larger and progressively incommensurate with $\lambda$, its influence on the SP weakens \cite{footnote2}. 

The above observation also provides a possible explanation for why the surface corrugations, commonly present in GaAs (001) samples, do not influence the SP \cite{Cooper.SSC.2001,Willett.PRL.2001}. Typically, SP's orientation is either fixed (hard-axis along $[\overline{1}10]$) or determined by $B_{||}$'s direction. In contrast, the alignment of the corrugations is sample-dependent and can occur primarily along [110] or $[\overline{1}10]$ directions \cite{Cooper.SSC.2001,Willett.PRL.2001}. The corresponding periods, as estimated from atomic force microscopy, are usually large compared to that of SPs \cite{Cooper.SSC.2001,Willett.PRL.2001}. While surface corrugations can induce anisotropic transport at $B=0$, they have no effect on the SPs. The likely reason, in view of Fig. 5 results, is therefore the mismatch of the respective length scales.

We conclude that even an extremely gentle ($<0.25\%$) external density modulation is sufficient to reorient the $B_{||}$-induced SP if the corresponding periods are comparable. Our technique to probe the structure and energetics of a SP should find use in studies of SPs in other strongly correlated systems. 

\begin{acknowledgments}
We acknowledge support through the NSF (Grants DMR-1305691 and ECCS-1508925) for measurements. We also acknowledge the NSF (Grant MRSEC DMR-1420541), the DOE BES (Grant DE-FG02-00-ER45841), the Gordon and Betty Moore Foundation (Grant GBMF4420), and the Keck Foundation for sample fabrication and characterization. Our measurements were partly performed at the National High Magnetic Field Laboratory (NHMFL), which is supported by the NSF Cooperative Agreement DMR-1157490, by the State of Florida, and by the DOE. We thank S. Hannahs, T. Murphy, A. Suslov, J. Park and G. Jones at NHMFL for technical support.
\end{acknowledgments}


\begin{thebibliography}{99}
\bibitem{footnote1} In real samples, the SP is nematic, i.e. imperfectly ordered because of disorder and thermal fluctuations at finite $T$. It is therefore reasonable to expect that, when in parallel, the imposed density modulation, despite being much weaker compared to the SP's intrinsic modulation, can strengthen the long range ordering of the SP.
\bibitem{footnote2} Note that $a=100$ nm is closer to the SP's intrinsic period compared to $a=175$ nm, but its modulation amplitude is smaller (see Fig. 4). This may explain why the relative resistivity increase is larger for $a=175$ nm.
\end{thebibliography}
\end{document}